\documentclass[emulateapj,twocolumn,
floatfix,preprintnumbers,altaffilletter]{revtex4}
\usepackage{graphicx,url,amssymb,amsmath,longtable,rotating,color,units,
wasysym,subfigure,epsfig}

\newcommand{\strain}{strain}
\begin{document}

\title{Directional limits on persistent gravitational waves using LIGO S5 science data}

\author{J.~Abadie$^{29}$, 
B.~P.~Abbott$^{29}$, 
R.~Abbott$^{29}$, 
M.~Abernathy$^{66}$, 
T.~Accadia$^{27}$, 
F.~Acernese$^{19ac}$, 
C.~Adams$^{31}$, 
R.~Adhikari$^{29}$, 
P.~Ajith$^{29}$, 
B.~Allen$^{2,78}$
% ...
}
% ...
\address{$^{2}$Albert-Einstein-Institut, Max-Planck-Institut f\"ur Gravitationsphysik, D-30167 Hannover, Germany}
% ...
\address{$^{19}$INFN, Sezione di Napoli $^a$; Universit\`a di Napoli 'Federico II'$^b$ Complesso Universitario di Monte S.Angelo, I-80126 Napoli; Universit\`a di Salerno, Fisciano, I-84084 Salerno$^c$, Italy}
% ...
\address{$^{27}$Laboratoire d'Annecy-le-Vieux de Physique des Particules (LAPP), Universit\'e de Savoie, CNRS/IN2P3, F-74941 Annecy-Le-Vieux, France}
% ...
\address{$^{29}$LIGO - California Institute of Technology, Pasadena, CA  91125, USA }
% ...
\address{$^{31}$LIGO - Livingston Observatory, Livingston, LA  70754, USA }
% ...
\address{$^{66}$University of Glasgow, Glasgow, G12 8QQ, United Kingdom }
% ...
\address{$^{78}$University of Wisconsin--Milwaukee, Milwaukee, WI  53201, USA }
% ...

\noaffiliation

\date{\today}

\begin{abstract}
The gravitational-wave (GW) sky may include nearby pointlike sources as well as astrophysical and cosmological stochastic backgrounds.
Since the relative strength and angular distribution of the many possible
sources of GWs are not well constrained, searches for GW signals must be performed in a model-independent way.
To that end we perform two directional searches for persistent GWs using data 
from the LIGO S5 science run:
one optimized for pointlike sources and one for arbitrary extended sources.
The latter result is the first of its kind.
Finding no evidence to support the detection of GWs, we present 90\% confidence level (CL)
upper-limit maps of GW strain power with typical values between
$\unit[2-20\times10^{-50}]{\strain^2 Hz^{-1}}$ and 
$\unit[5-35\times10^{-49}]{\strain^2 Hz^{-1}sr^{-1}}$ 
for pointlike and extended sources respectively.
The limits on pointlike sources constitute a factor of $30$ improvement over the previous best limits.
We also set 90\% CL limits on the narrow-band root-mean-square GW strain from interesting targets including Sco X-1, SN1987A and the Galactic Center as low as $\approx7\times10^{-25}$ in the most sensitive frequency range near $\unit[160]{Hz}$.
These limits are the most constraining to date and constitute a factor of $5$ improvement over the previous best limits.
\end{abstract}

\maketitle

\section{Introduction}\label{intro}
One of the most ambitious goals of gravitational-wave (GW) astronomy is to measure the stochastic cosmological gravitational-wave background (CGB), which can arise through a variety of mechanisms including amplification of vacuum fluctuations following inflation~\cite{kolbturner}, phase transitions in the early universe~\cite{starobinksii,bar-kana}, cosmic strings~\cite{kibble,damour} and pre-Big Bang models~\cite{buonanno,mandic}.
The CGB may be masked by an astrophysical gravitational-wave background (AGB), interesting in its own right, which can arise from the superposition of unresolved sources such as core-collapse supernovae~\cite{howell,ferrari}, neutron-star excitations~\cite{ferrari2,sigl}, binary mergers~\cite{tania_cbc2,farmer}, persistent emission from neutron stars~\cite{tania_ns,tania_ns2} and compact objects around supermassive black holes~\cite{barack,sigl2}.

We present the results of two analyses using data from the LIGO S5 science run:
a radiometer analysis optimized for pointlike 
sources and a spherical-harmonic decomposition analysis, 
which allows for arbitrary angular distributions.
This work presents the first measurement of the GW sky in a framework consistent with an arbitrary extended source.

\section{LIGO Detectors and the S5 Science Run}\label{S5}
We analyze data from LIGO's $\unit[4]{km}$ and $\unit[2]{km}$ detectors 
(H1 and H2) in Hanford, WA and the $\unit[4]{km}$ detector 
(L1) in  Livingston Parish, LA during the S5 science run, 
which took place between Nov.~5, 2005 and Sep.~30, 2007.
During S5, both H1 and L1 reached a strain sensitivity of $\unit[3\times10^{-23}]{strain\,Hz^{-1/2}}$ in the most sensitive region between $\unit[100-200]{Hz}$~\cite{S5} and collected $\unit[331]{days}$ of coincident H1L1 and H2L1 data.
S5 saw milestones in GW astronomy including 
limits on the emission of GWs from the Crab pulsar 
that surpass those inferred from the Crab's spindown~\cite{crab-S5},
as well as limits on the isotropic CGB that surpass indirect 
limits from Big Bang nucleosynthesis and the cosmic microwave background~\cite{stoch-S5}.
This work builds on~\cite{stoch-S5,radiometer}.

\section{Methodology}\label{methods}
Following~\cite{sph_methods,radiometer} we present a framework for analyzing 
the angular distribution of GWs.
We assume that the GW signal is stationary 
and unpolarized, but not necessarily isotropic.
It follows that the GW energy density  $\Omega_\text{GW}(f)$,
can be expressed in terms of the GW power spectrum, 
${\cal P}(f,\hat\Omega)$:
\begin{equation}
  \Omega_\text{GW}(f)\equiv\frac{f}{\rho_c}\frac{d\rho_\text{GW}}{df}
  =\frac{2\pi^2}{3H_0^2}{f^3} \int_{S^2}d\hat\Omega\>{\cal P}(f,\hat\Omega) .
  \label{e:OmegaGW}
\end{equation}
Here $f$ is frequency, $\hat\Omega$ is sky location, $\rho_c$ is the critical density of the universe 
and $H_0$ is Hubble's constant.
We further assume that 
${\cal P}(f,\hat\Omega)$
can be factored 
(in our analysis band) into an angular power spectrum, 
${\cal P}(\hat{\Omega})$, and a spectral shape,
$\bar{H}(f)\equiv(f/f_0)^\beta$, parameterized by 
the spectral index $\beta$ and reference frequency $f_0$.
We set $f_0=\unit[100]{Hz}$ to be in the sensitive range of the LIGO interferometers.

Our goal is to 
measure ${\cal P}(\hat{\Omega})$ for two power-law signal models.
In the cosmological model, 
$\beta=-3$ ($\Omega_\text{GW}(f)=\text{const}$), which is predicted, e.g., for the amplification of vacuum fluctuations following inflation~(see \cite{maggiore} and references therein).
In the astrophysical model, $\beta=0$ ($\bar H(f)=\text{const}$), which emphasizes the strain sensitivity of the LIGO detectors.

We estimate ${\cal P}(\hat{\Omega})$ two ways.
The {\em radiometer algorithm}~\cite{radiometer,radio_method,radio_method2}
assumes the signal is a point source characterized by a single direction $\hat\Omega_0$ and amplitude, $\eta(\hat\Omega_0)$:
\begin{equation}\label{eq:eta}
  {\cal P}(\hat\Omega) \equiv 
  \eta(\hat\Omega_0) \,\delta^2(\hat\Omega,\hat\Omega_0) .
\end{equation}
It is applicable to a GW sky dominated by a limited number of widely separated point sources.
As the number of point sources is increased, however, the interferometer beam pattern will cause the signals to interfere and partly cancel.
Thus, radiometer maps do not apply to extended sources.
Since pointlike signals are expected to arise from astrophysical sources, we use $\beta=0$ for the radiometer analysis.

The {\em spherical-harmonic decomposition (SHD) algorithm} is
used for both $\beta=-3$ (cosmological) and $\beta=0$ (astrophysical) sources.
It allows for the possibility of an extended source with an 
arbitrary angular distribution, characterized by 
spherical-harmonic coefficients ${\cal P}_{lm}$ such that
\begin{equation} 
  {\cal P}(\hat{\Omega})\equiv\sum_{lm}{\cal P}_{lm}Y_{lm}(\hat\Omega) .
\end{equation}
The series is cut off at $l_\text{max}$, allowing for angular scale $\sim2\pi/l_\text{max}$.
The flexibility of the spherical-harmonic algorithm comes 
at the price of somewhat diminished sensitivity to point sources, 
and thus the two algorithms are complementary.

We choose $l_\text{max}$ so as to minimize 
the sky average
of the product of $\sigma(\hat\Omega)$ and $\bar A$,
where $\sigma(\hat\Omega)$ is the uncertainty associated with 
${\cal P}(\hat\Omega)$  and $\bar A$ is the typical angular
area of a resolved patch of sky~\footnote{The data were first processed with $l_\text{max}=20$; the present method of choosing $l_\text{max}$ was then adopted a posteriori in order to more accurately model the angular resolution of the interferometer network.}.
Since $\bar{A}={4\pi}/N_\text{indep}\propto 1/(l_\text{max}+1)^2$ 
(where $N_\text{indep}$ is the number of independent parameters),
this procedure amounts to choosing 
$l_\text{max}$ to maximize the sensitivity obtained 
by integrating over the typical search aperture (angular resolution).
We obtain $l_\text{max}=7$ and $12$ for 
$\beta=-3$ and $\beta=0$, respectively.
Since the search aperture becomes smaller at the higher frequencies emphasized by $\beta=0$, $l_\text{max}$ is larger for $\beta=0$ than for $\beta=-3$.

Both algorithms can be framed in terms of a ``dirty map'', 
$X_{\nu}$, which represents the signal convolved the Fisher matrix, $\Gamma_{\mu\nu}$:
\begin{eqnarray}
  X_{\nu} = \sum_{ft} \gamma_{\nu}^\star (f,t) \frac{\bar{H}(f)}{P_1(f,t)P_2(f,t)}C(f,t) \\
  \Gamma_{\mu\nu} = \sum_{ft} \gamma_{\mu}^\star(f,t) 
  \frac{\bar{H}^2(f)}{P_1(f,t)P_2(f,t)}\gamma_\nu(f,t)\,.
\end{eqnarray}
Here both the Greek indices $\mu$ and $\nu$ take on values of $lm$ for the SHD algorithm and $\hat\Omega$ for the radiometer algorithm, for which we use the pixel basis.
The two bases are related using spherical-harmonic basis functions:
\begin{equation}
  X_{\hat\Omega} = \sum_{lm} X_{lm} Y_{lm}(\hat\Omega) .
\end{equation}
$C(f,t)$, meanwhile, is the cross spectral density generated from the H1L1 or H2L1 pairs.
$P_1(f,t)$ and $P_2(f,t)$ are the individual power spectral densities,
and $\gamma_\mu(f,t)$ is the angular decomposition of the overlap reduction 
function $\gamma(\hat\Omega,f,t)$, which characterizes the orientations and frequency response of the detectors~\cite{sph_methods}:
\begin{eqnarray}
  \gamma_{\mu}(f, t)
  & \equiv & \int_{S^{2}} d\hat\Omega \,
  \gamma(\hat{\Omega}, f, t)\, {\bf e}_\mu(\hat\Omega)
  \\
  \gamma({\hat\Omega},f,t) & = &
  \frac{1}{2} F_1^A(\hat\Omega,t)F_2^A(\hat\Omega,t)
  e^{i2\pi f\hat\Omega\cdot(\Delta\vec{x}_{12}(t))/c} .
  \label{eq:gamma}
\end{eqnarray}
Here $F_I^A(\hat\Omega,t)$ characterizes the detector response of detector $I$ to a GW with polarization $A$, ${\bf e}_\mu(\hat\Omega)$ is a basis function, $c$ is the speed of light and $\Delta\vec{x}_{12}\equiv\vec{x}_1-\vec{x}_2$ is the difference between the interferometer locations.
A detailed discussion of these quantities can be found in~\cite{sph_methods}.

In~\cite{sph_methods} it was shown that the maximum-likelihood estimators
of GW power are given by $\hat{\cal P}=\Gamma^{-1}X$.
The inversion of $\Gamma$ is complicated
by singular eigenvalues 
associated with modes to which the Hanford-Livingston (HL) detector network is insensitive.
This singularity can be handled two ways.
The radiometer algorithm assumes the signal is pointlike, 
implying that correlations between neighboring pixels can be ignored.
Consequently, we can replace $\Gamma^{-1}$ with 
$(\Gamma_{\hat\Omega\hat\Omega})^{-1}$ to estimate the point source amplitude $\eta(\hat\Omega)$ (see Eq.~\ref{eq:eta}).
(We note that pointlike sources create signatures in our sky maps that 
typically span several degrees or more; see~\cite{radiometer}.)

The SHD algorithm, on the other hand, targets extended sources, so the full Fisher matrix must be taken into account. 
We regularize $\Gamma$ 
by removing a fraction, $\cal F$, of the modes associated 
with the smallest eigenvalues, to which the HL network is relatively insensitive.
$\cal F$ is known as the regularization cutoff.
By removing some modes from the Fisher matrix, we obtain a 
regularized inverse Fisher matrix, 
$\Gamma_R^{-1}$, thereby introducing a bias, 
the implications of which are discussed below.
For now, we note that the bias depends on the angular distribution of the signal.

We thereby obtain the estimators
\begin{eqnarray}\label{eq:estimators}
  \hat{\eta}_{\hat\Omega} & = & 
  \left(\Gamma_{\hat\Omega\hat\Omega}\right)^{-1} X_{\hat\Omega} 
  \label{eq:p_rad}
  \\
  \hat{\cal P}_{lm} & = & \sum_{l'm'}(\Gamma_R^{-1})_{lm,l'm'} X_{l'm'} \,,
\end{eqnarray}
with uncertainties 
\begin{align}
  \sigma^\text{rad}_{\hat\Omega}&=(\Gamma_{\hat\Omega\hat\Omega})^{-1/2}
  \label{e:sigma_rad}
  \\
  \sigma^\text{sph}_{lm}&=\left[(\Gamma^{-1}_R)_{lm,lm}\right]^{1/2}\,.
  \label{e:sigma_sph}
\end{align}
We refer to $\hat{\cal P}_{\hat\Omega}\equiv\sum_{lm}\hat{\cal P}_{lm}Y_{lm}(\hat\Omega)$ as the ``clean map" and $\hat\eta_{\hat\Omega}$ as the ``radiometer map.''
We note that $\hat\eta_{\hat\Omega}$ has units of $\unit[]{strain^2Hz^{-1}}$ whereas $\hat{\cal P}_{\hat\Omega}$ has units of $\unit[]{strain^2Hz^{-1}sr^{-1}}$.

In choosing $\cal F$ one must balance the competing demands of reconstruction accuracy (sensitivity to the modes that are kept) with the bias associated with the modes that are removed.
In practice, we do not know the bias associated with $\cal F$ since it depends on the unknown signal distribution ${\cal P}(\hat\Omega)$.
Therefore, we choose a value of $\cal F$
that tends to produce reliably reconstructed maps with minimal bias for simulated signals.
Following~\cite{sph_methods}, we use ${\cal F}=1/3$, which was shown to be a robust regularization cutoff for simulated signals including maps characterized by one or 
more point sources, dipoles, monopoles and an extended source 
clustered in the galactic plane (see~\cite{sph_methods}).

In the case of the SHD algorithm, we construct an additional statistic (see~\cite{sph_methods}),
\begin{equation}
  \hat C_l \equiv  \frac{1}{2l+1} \sum_m \left[ 
    |\hat{\cal P}_{lm}|^2 - (\Gamma_R^{-1})_{lm,lm } \right] \,,
\end{equation}
which describes the angular scale of the clean map.
The subtracted second term makes the estimator unbiased
so that $\langle\hat{C}_l\rangle=0$ when no signal is present.
The expected noise distribution of $\hat C_l$ is highly non-Gaussian 
for small values of $l$, and so the 
upper limits presented below are calculated numerically.
The $\hat C_l$ are analogous to 
similar quantities defined in the context of temperature fluctuations of the cosmic microwave background (see, e.g.,~\cite{wmap}).

The analysis was performed using the S5 stochastic analysis pipeline.
This pipeline has been tested with hardware and software injections, and the 
successful recovery of isotropic hardware injections is documented in~\cite{stoch-S5}.
The recovery of anisotropic software injections is demonstrated in~\cite{sph_methods}.
We parse time series into $\unit[60]{s}$, Hann-windowed, 50\%-overlapping segments, which are coarse-grained to achieve $\unit[0.25]{Hz}$ resolution.
We apply a stationarity cut described in~\cite{radiometer}, 
which rejects $\sim3\%$ of the cross-correlated segments.
We also mask frequency bins associated with instrumental lines 
(e.g., harmonics of the 60~Hz mains power, calibration lines 
and suspension-wire resonances) as well as injected, simulated pulsar signals.
For $\beta=-3,0$ we include frequency bins up to $\unit[200,500]{Hz}$, so that $\sigma({\hat\Omega})$ 
is within $\lesssim2\%$ of the minimum possible value.
Thirty-three frequency bins are masked, corresponding to 
$2\%$ of the frequency bins between $\unit[40-500]{Hz}$ used 
in the broadband analyses.
For additional details about the S5 stochastic pipeline, see~\cite{stoch-S5}.

\section{Significance and upper limit calculations}\label{stats}
In order to determine if there is a statistically significant GW signature, 
we are primarily interested in the significance of outliers---the highest signal-to-noise ratio ($\text{SNR}$) frequency bin or sky-map pixel.
It is therefore necessary to calculate the expected noise 
probability distribution of the maximum $\text{SNR}$ given 
many independent trials (when considering maximum 
$\text{SNR}$ in a spectral band) and given many {\em dependent} 
trials (when considering maximum $\text{SNR}$ for a sky map).

For $N$ independent frequency bins, 
the probability density function, $\pi(\rho_\text{max})$, 
of maximum $\text{SNR}$, $\rho_\text{max}$, is given by
\begin{equation}\label{eq:max_snr}
  \pi(\rho_\text{max}) \propto \left[1+\text{erf}\left(\rho_\text{max}/\sqrt{2}\right) 
  \right]^{N-1}
  e^{-\rho_\text{max}^2/2}\, .
\end{equation}
Here we have assumed that the stochastic point estimate is Gaussian distributed.
The Gaussianity of $\hat{\cal P}_{\hat\Omega}$ and $\hat\eta_{\hat\Omega}$, calculated by summing over many ${\cal O}(500\text{K})$ independent segments, is expected to arise due to the central limit theorem~\cite{allen-romano}.
Additionally, we find the Gaussian-noise hypothesis to be consistent with 
time-slide studies, wherein we perform the cross-correlation analysis with an unphysical time-shift in order to wash out astrophysical signals and thereby obtain different realizations of detector noise.

\begin{figure}[hbtp!]
      \psfig{file=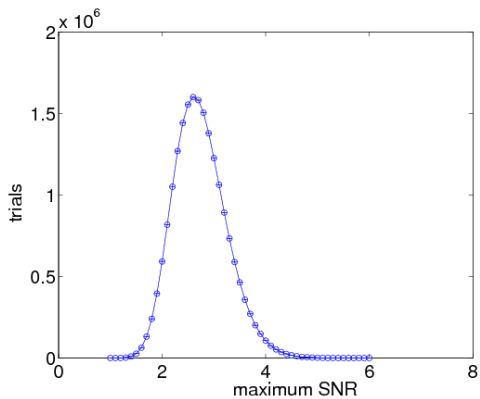,width=2.5in}
      \caption{Numerically calculated  distribution of the maximum $\text{SNR}$ for $\beta=-3$ clean maps created from Gaussian noise. \label{fig:maptest}}
\end{figure}

The distribution of maximum SNR for a sky map is more subtle
due to the non-zero covariances that exist between different
pixels (or patches) on the sky.
For this case, we calculate $\pi(\rho_\text{max})$ numerically,
by simulating many realizations of dirty maps that have 
expected covariances described by the Fisher matrix $\Gamma$.
Figure~\ref{fig:maptest} shows the numerically determined 
$\pi(\rho_\text{max})$ for the $\beta=-3$ clean map 
generated with Gaussian noise.

The likelihood function for ${\cal P}(\hat\Omega)$ 
at each point in the sky 
can be be described as a normal distribution with mean 
$\hat{\cal P}_{\hat\Omega}$ and a variance $(\sigma^\text{sph}_{\hat\Omega})^2$.
In the case of the SHD algorithm, regularization introduces a signal-dependent bias.
Without knowing the true distribution of ${\cal P}(\hat\Omega)$, 
it is impossible to know the bias exactly, but it is 
possible to set a conservative upper limit by assuming 
that on average the modes removed through regularization contain 
no more GW power than the modes that are kept.

To implement this assumption, we calculate 
$\hat{\cal P}_{lm}$ with a regularization scheme that 
sets eigenvalues of removed modes to zero, 
whereas $\sigma_{lm}^\text{sph}$ is conservatively calculated using a 
regularization scheme that sets eigenvalues of removed modes 
to the average eigenvalue of the kept modes.
This has the effect of widening the likelihood function at each 
sky location.
The upper limits become on average $25\%$ larger than 
they would be if we had calculated $\sigma_{lm}^\text{sph}$ 
using the same regularization scheme as $\hat{\cal P}_{lm}$.

Following the same procedure as in~\cite{stoch-S5},
we marginalize over the H1, H2, and L1 
calibration uncertainties, which were measured to be 10\%,
10\%, and 13\%, respectively~\cite{S5calib}~\footnote{We follow the marginalization scheme used in~\cite{stoch-S5}, but note that this does not take into account covariance between baselines with a shared detector.  Because the H2L1 baseline only contributes about 10\% of the sensitivity, and the calibration uncertainty is on the order of 10\% to start with, this effect is only on the order of 1\%.  Work is ongoing to take this effect into account, which we expect to be more important for baselines with comparable sensitivities.}
The posterior distribution is obtained by multiplying the 
marginalized likelihood function by a prior, which we take 
to be flat above ${\cal P}(\hat\Omega)>0$~\footnote{A prior constructed from~\cite{radiometer} would be nearly flat anyway since the strain sensitivity has improved ten-fold since the S4 science run.}.
The Bayesian upper limits are then determined by integrating 
the posterior out to the value of ${\cal P}(\hat\Omega)$ 
which includes 90\% of the total area under the distribution.
The calculation of upper limits on $\eta_{\hat\Omega}$ is analogous except we need not take into account the effects of regularization.

\section{Results}\label{results}
\begin{figure*}[t!]
  \begin{tabular}{ccc}
      \psfig{file=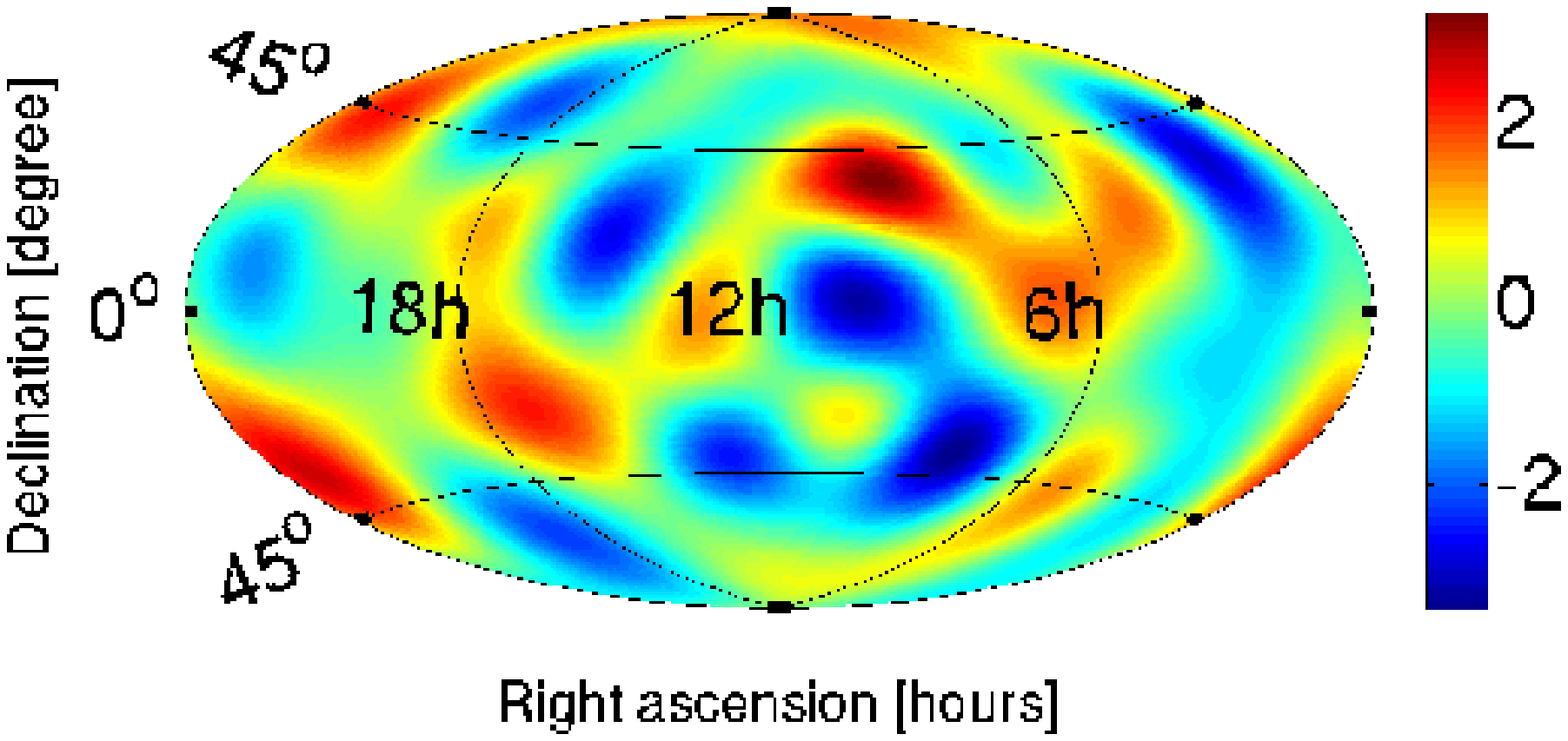,width=2.25in} & 
      \psfig{file=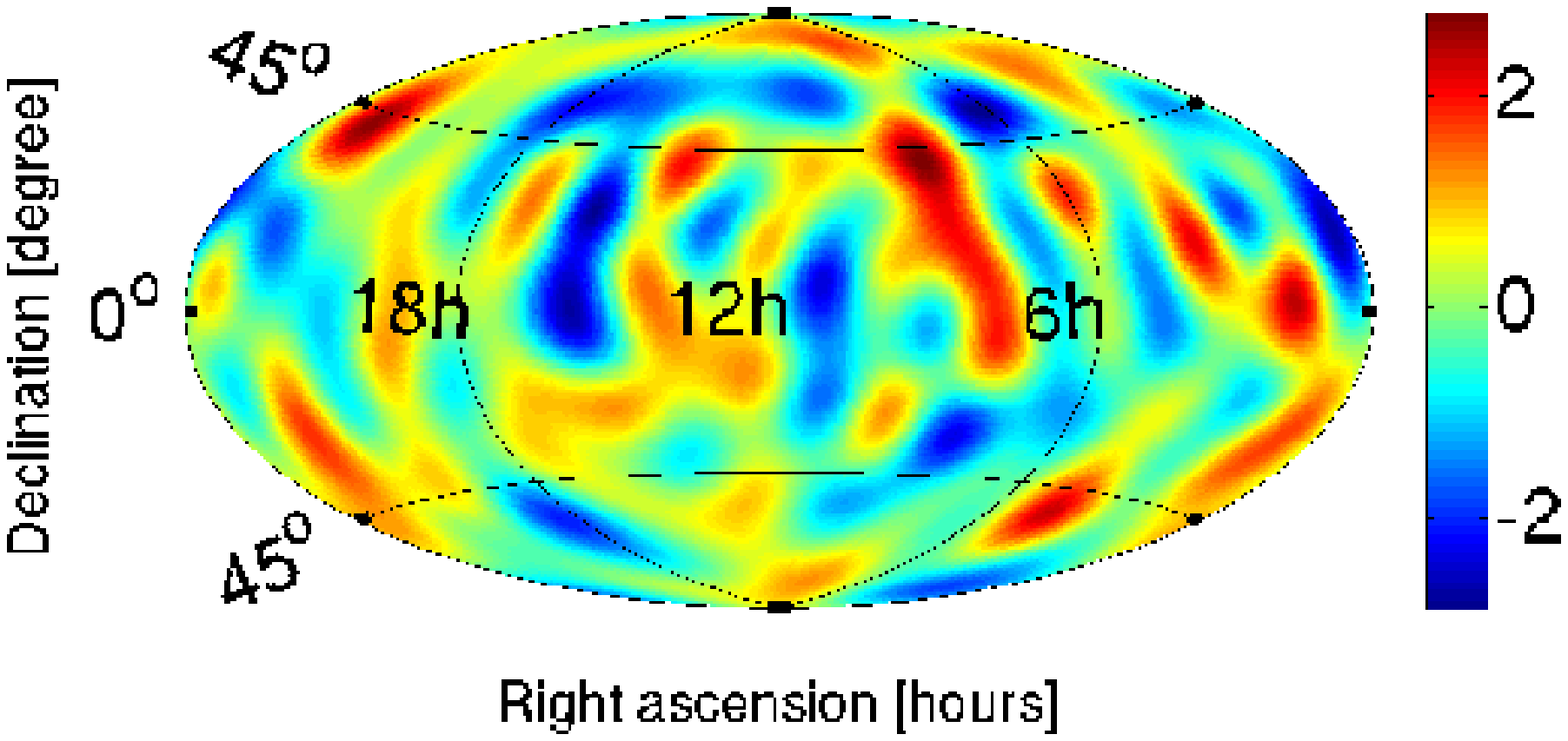,width=2.25in} & 
      \psfig{file=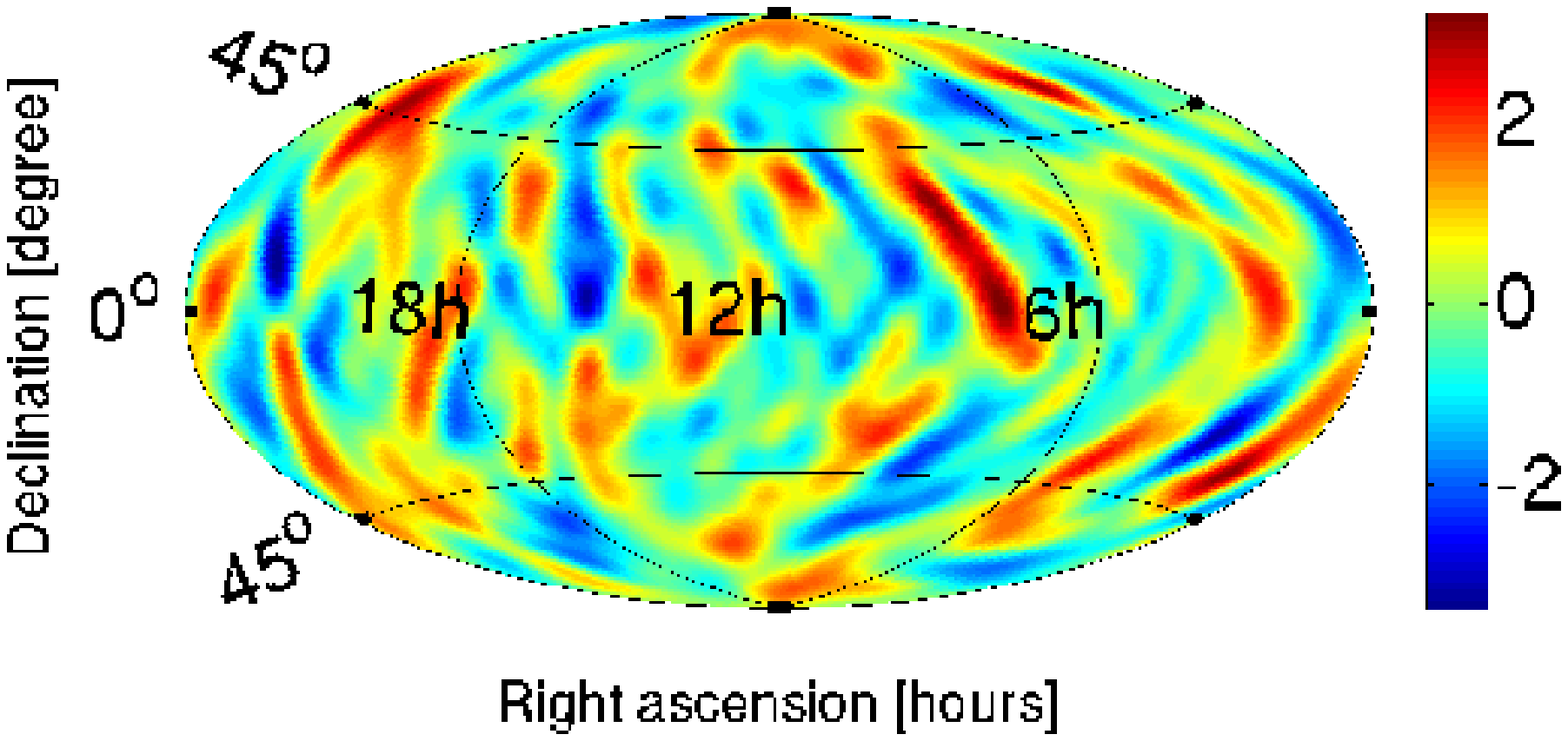,width=2.25in} \\ 
      \psfig{file=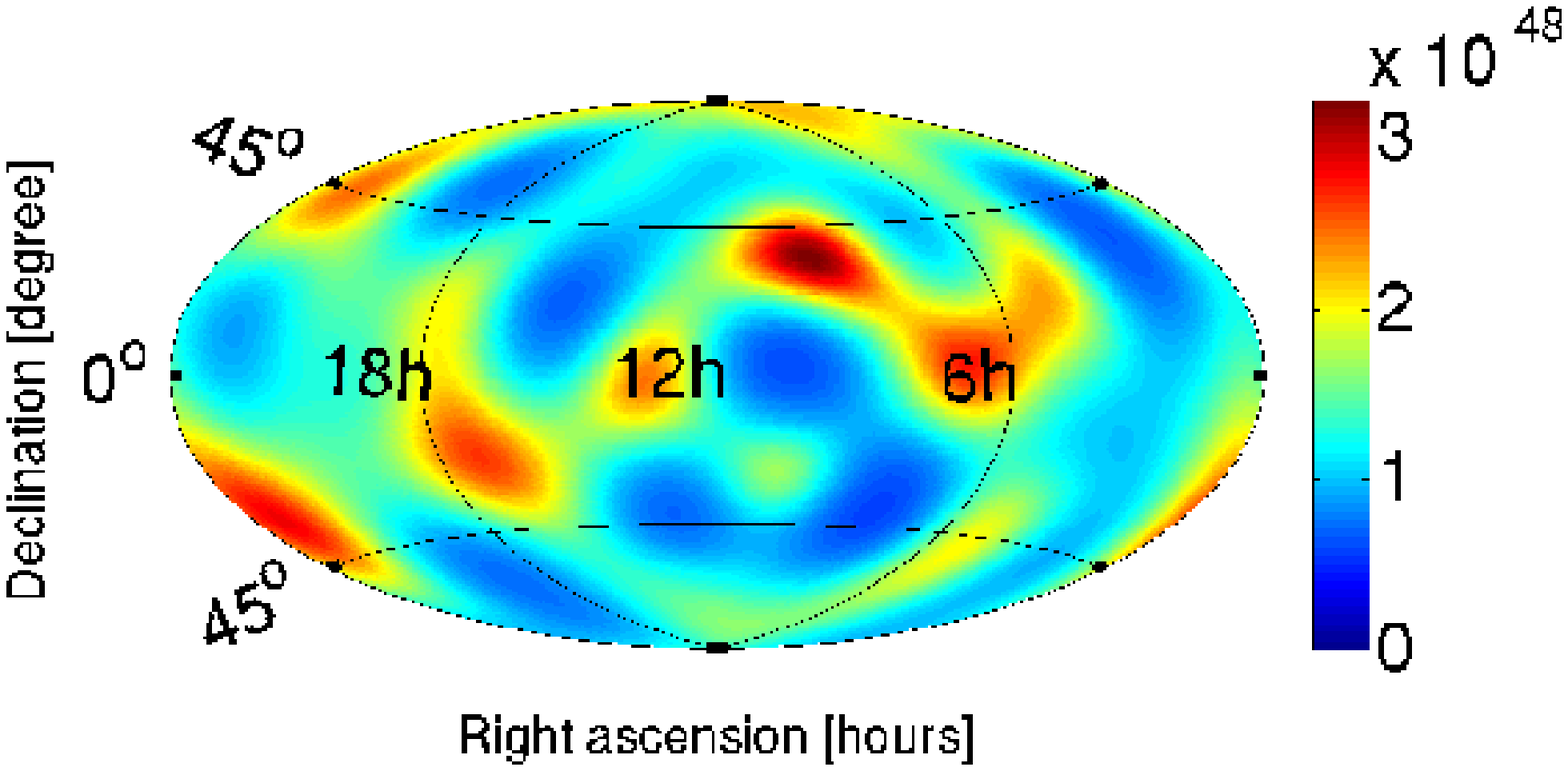,width=2.25in} &  
      \psfig{file=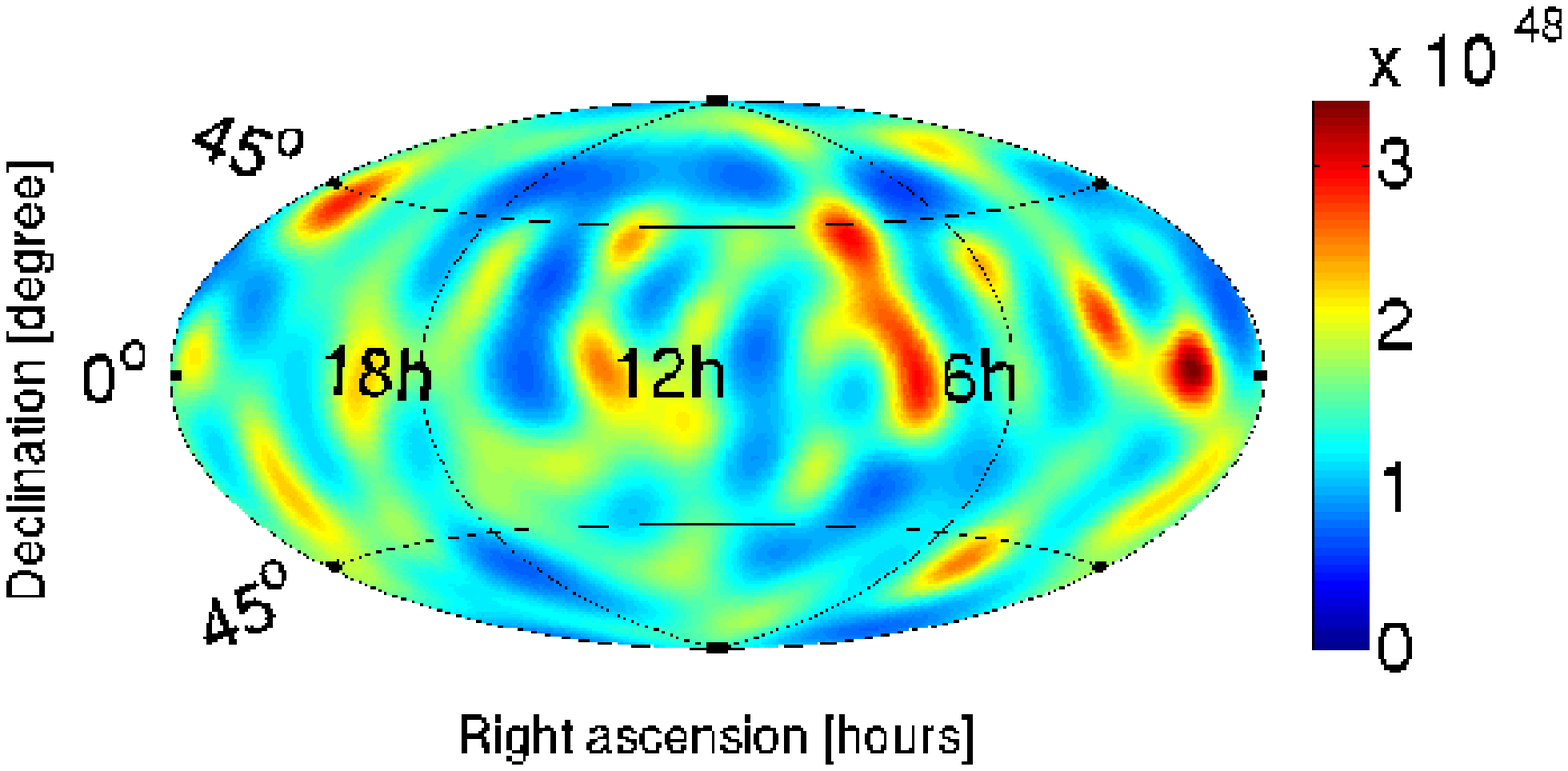,width=2.25in} & 
      \psfig{file=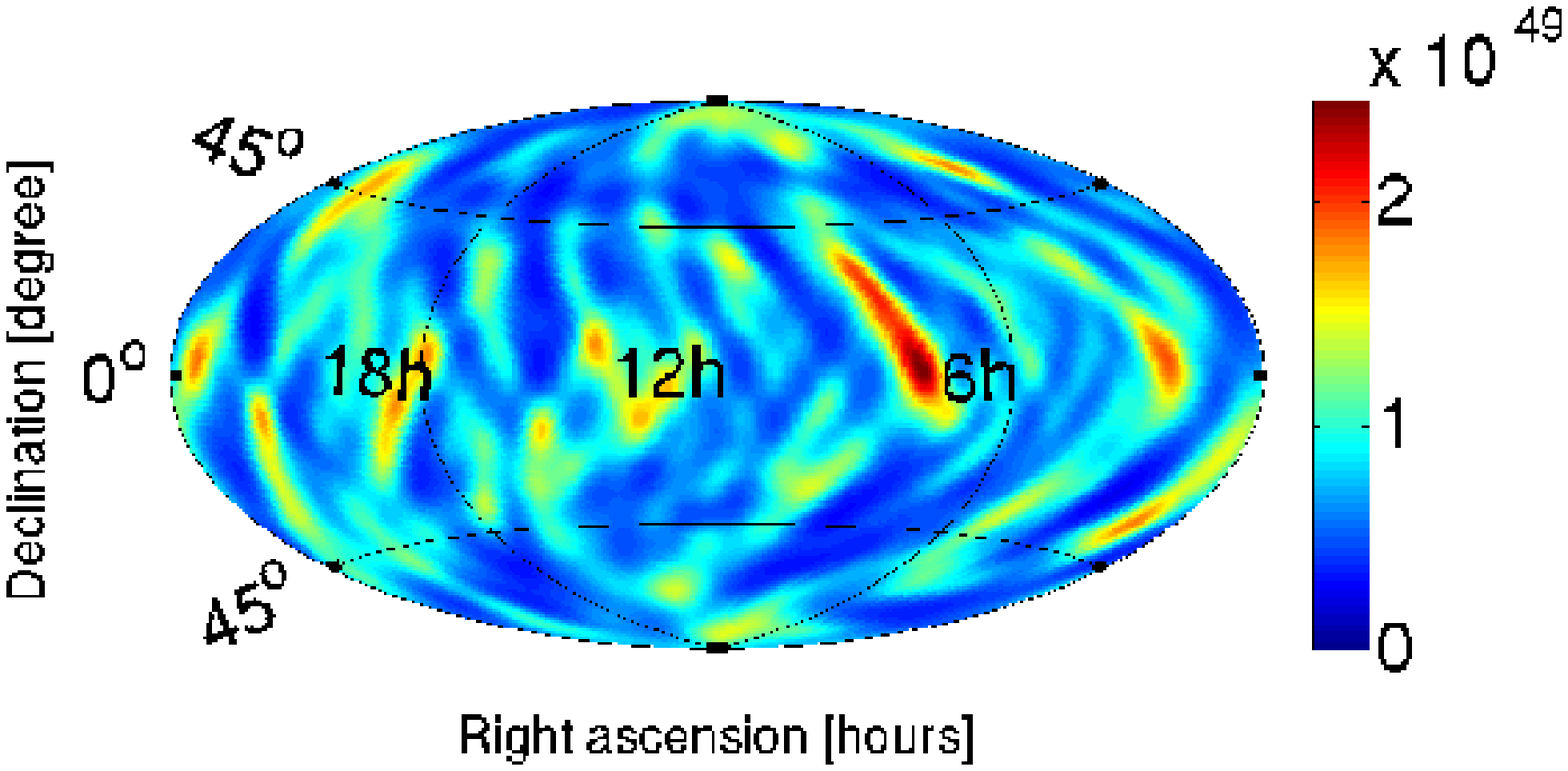,width=2.25in} \\ 
  \end{tabular}
  \caption{
    Top row: Signal-to-noise ratio maps for the three different analyses described
    in this paper:
    SHD clean map for $\beta=-3$ (left),
    SHD clean map for $\beta=0$ (center), and
    radiometer for $\beta=0$ (right).
    All three SNR maps are consistent with detector noise.
    The p-values associated with each map's maximum $\text{SNR}$ are (from left to right) $p=25\%,\,p=56\%,\,p=53\%$.
    Bottom row: The corresponding 90\% CL upper limit maps on strain power
    in units of 
    $\unit[]{\strain^2 Hz^{-1}sr^{-1}}$ for the SHD algorithm, and units of
    $\unit[]{\strain^2 Hz^{-1}}$ for the radiometer algorithm.
    \label{fig:skymaps}
  }
\end{figure*}

{\em Sky maps:}
Figure~\ref{fig:skymaps} presents sky map results for 
the different analyses:
SHD algorithm with $\beta=-3$ 
(left),
SHD with $\beta=0$
(center),
and radiometer with $\beta=0$
(right).
The top row contains SNR maps.
The maximum $\text{SNR}$ values are $3.1$ (with significance $p=25\%$), $3.1$ (with $p=56\%$), and $3.2$ (with $p=53\%$) respectively.
These $p$-values take into account the number of search directions and covariances between different sky patches (see~\ref{stats}).
Observing no evidence of GWs, we set upper limits on 
GW power as a function of direction.
The 90\% confidence level (CL) 
upper limit maps are given in the bottom row.
For the SHD method with $\beta=-3$,
the limits are between $\unit[5-31\times10^{-49}]{\strain^2 Hz^{-1}sr^{-1}}$;
for SHD with $\beta=0$, 
the limits are between $\unit[6-35\times10^{-49}]{\strain^2 Hz^{-1} sr^{-1}}$;
and for the radiometer with $\beta=0$,
the limits are between $\unit[2-20\times10^{-50}]{\strain^2 Hz^{-1}}$.

The strain power limits can also be expressed in terms of
the GW energy flux per unit frequency~\cite{radiometer}:
\begin{align}
  \hat F(f,\hat\Omega) & = \frac{c^3 \pi f_0^2}{4G} \left(\frac{f}{f_0}\right)^{\beta+2} \hat {\cal P}_{\hat\Omega} \\ 
  & = \left(\unit[3.18\times10^{42}]{\frac{erg}{cm^2 s}}\right) \left(\frac{f}{\unit[100]{Hz}}\right)^{\beta+2} \hat {\cal P}_{\hat\Omega} .
\end{align}
(Radiometer energy flux is obtained by replacing $\hat{\cal P}_{\hat\Omega}$ with $\hat\eta_{\hat\Omega}$.)
The corresponding values are
$\unit[2-10\times10^{-6}(f/\unit[100]{Hz})^{-1}]{erg\,cm^{-2}s^{-1}Hz^{-1}sr^{-1}}$ and $\unit[2-11\times10^{-6}(f/\unit[100]{Hz})^2]{erg\,cm^{-2}s^{-1}Hz^{-1}sr^{-1}}$ for the SHD method, and $\unit[6-60\times10^{-8}(f/\unit[100]{Hz})^2]{erg\,cm^{-2}s^{-1}Hz^{-1}}$ for the radiometer.
The radiometer map constitutes a factor of $\sim30$ improvement over the 
previous best strain power limits~\cite{radiometer}.

\begin{figure*}[htbp!]
  \begin{tabular}{cc}
      \psfig{file=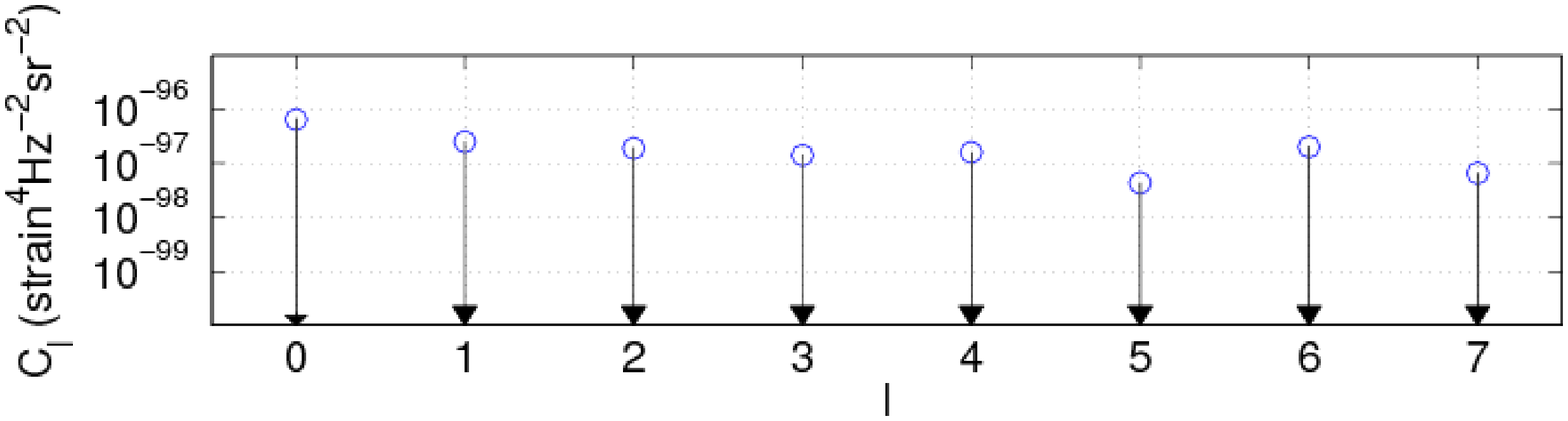,width=3in} & 
      \psfig{file=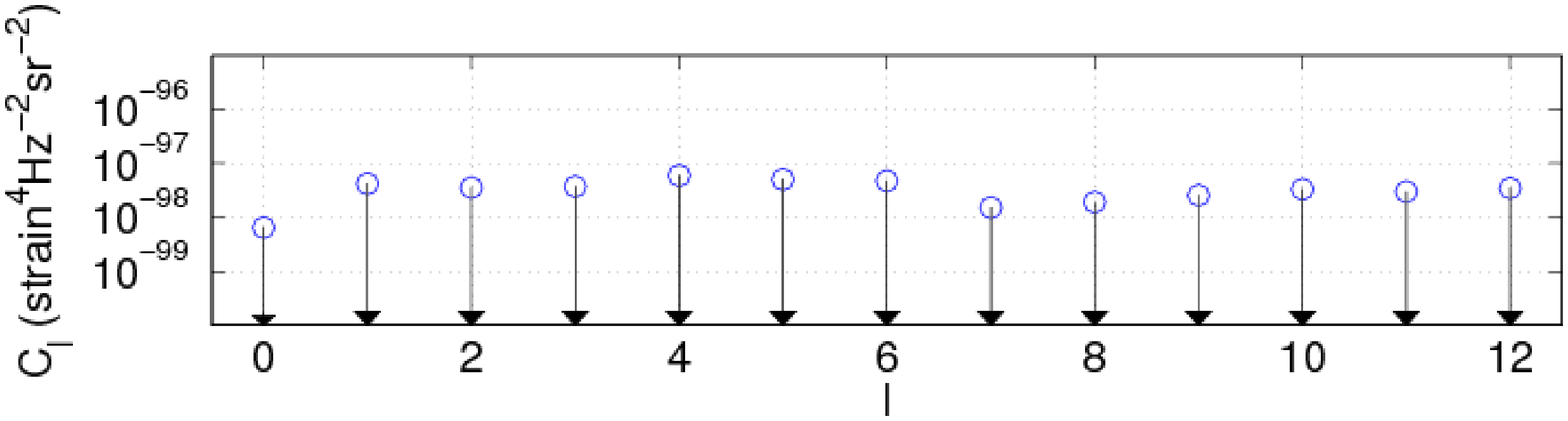,width=3in} \\ 
  \end{tabular}
  \caption{
    Upper limits on $C_l$ at 90\% CL vs $l$ 
    for the SHD analyses for $\beta=-3$ (left) and $\beta=0$ (right).
    The $\hat C_l$ are consistent with detector noise.
    \label{fig:Cls}
  }
\end{figure*}
\begin{figure*}[htbp!]
  \begin{tabular}{cc}
      \psfig{file=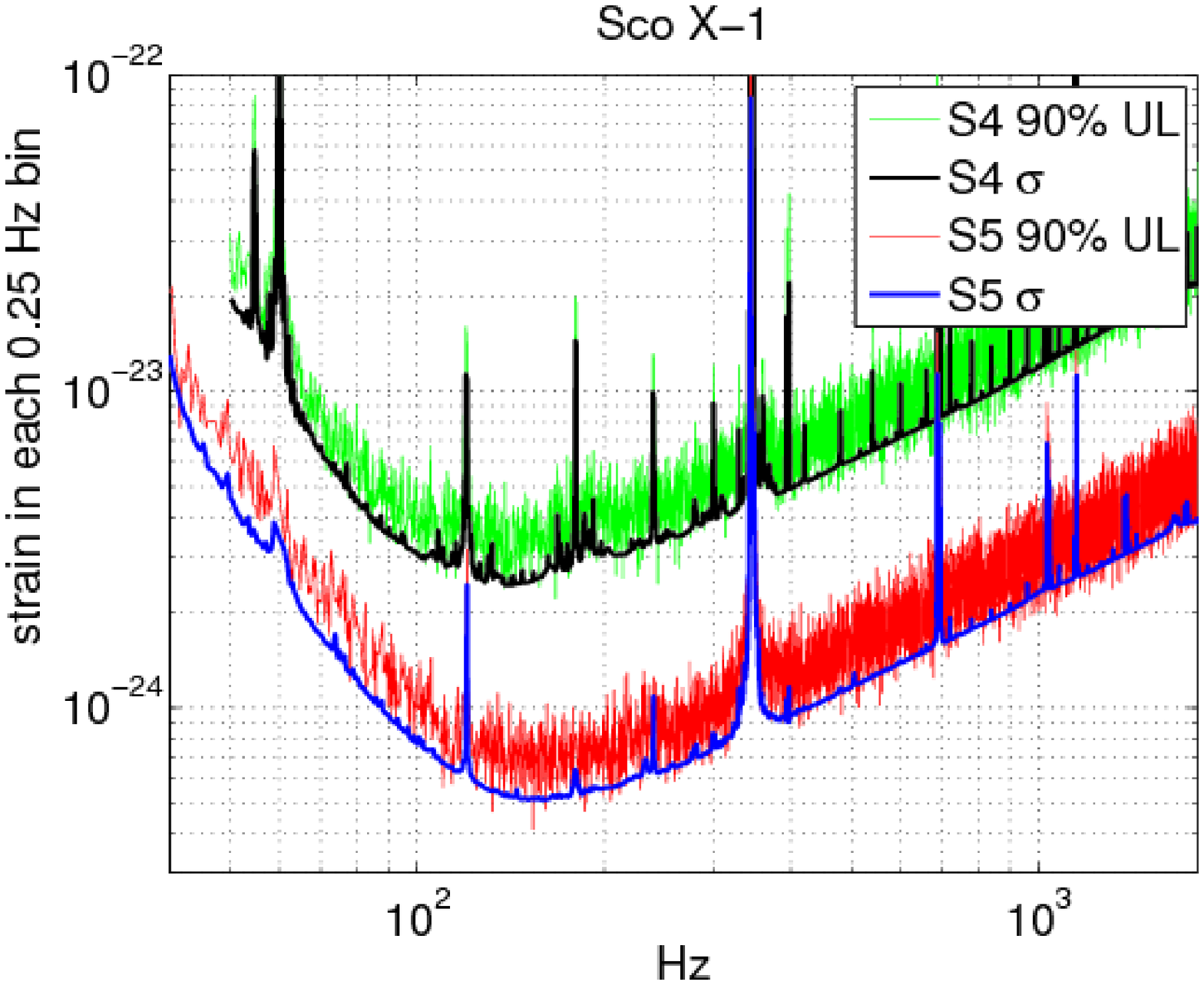,width=2.5in} & 
      \psfig{file=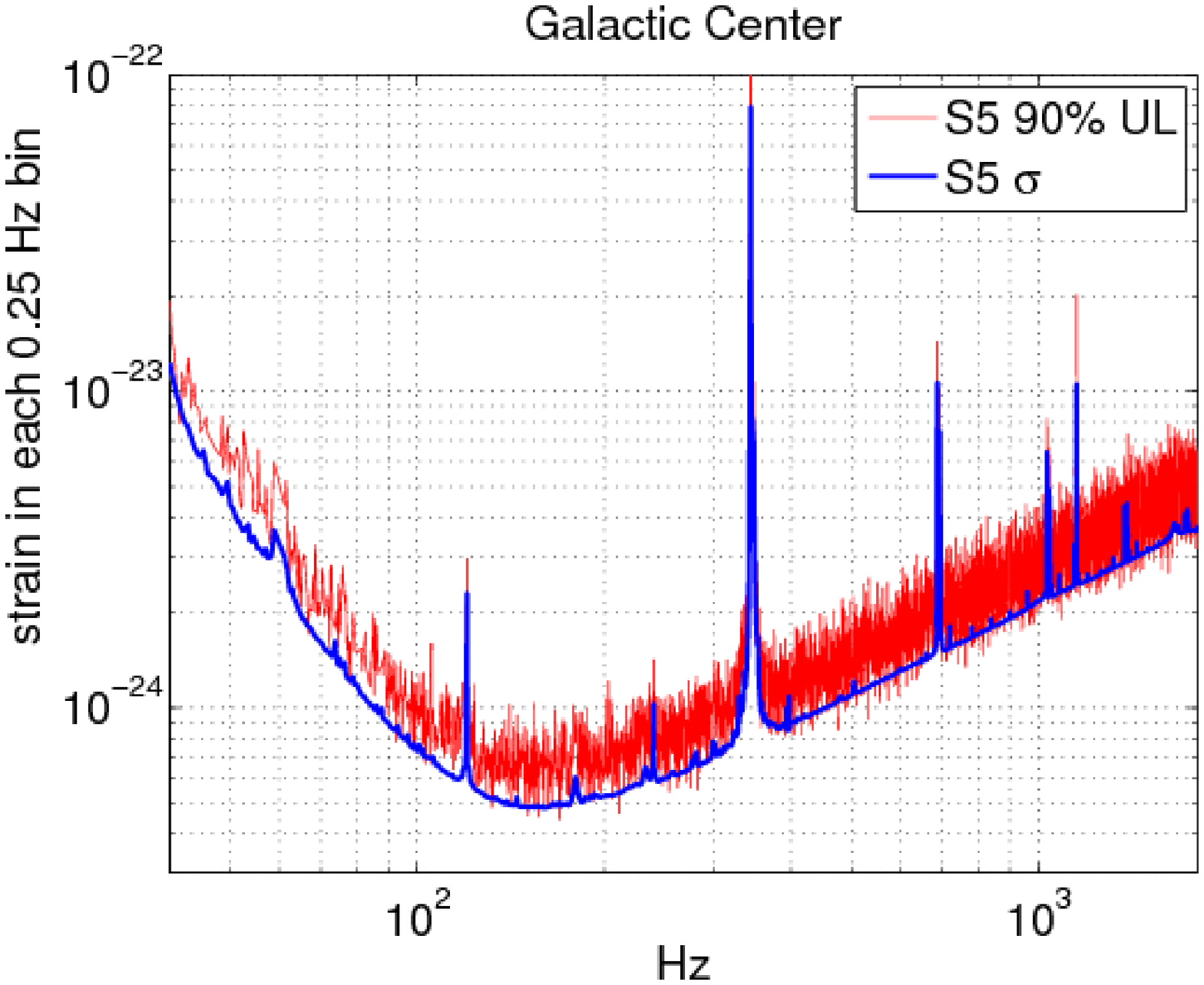,width=2.5in} \\
     \psfig{file=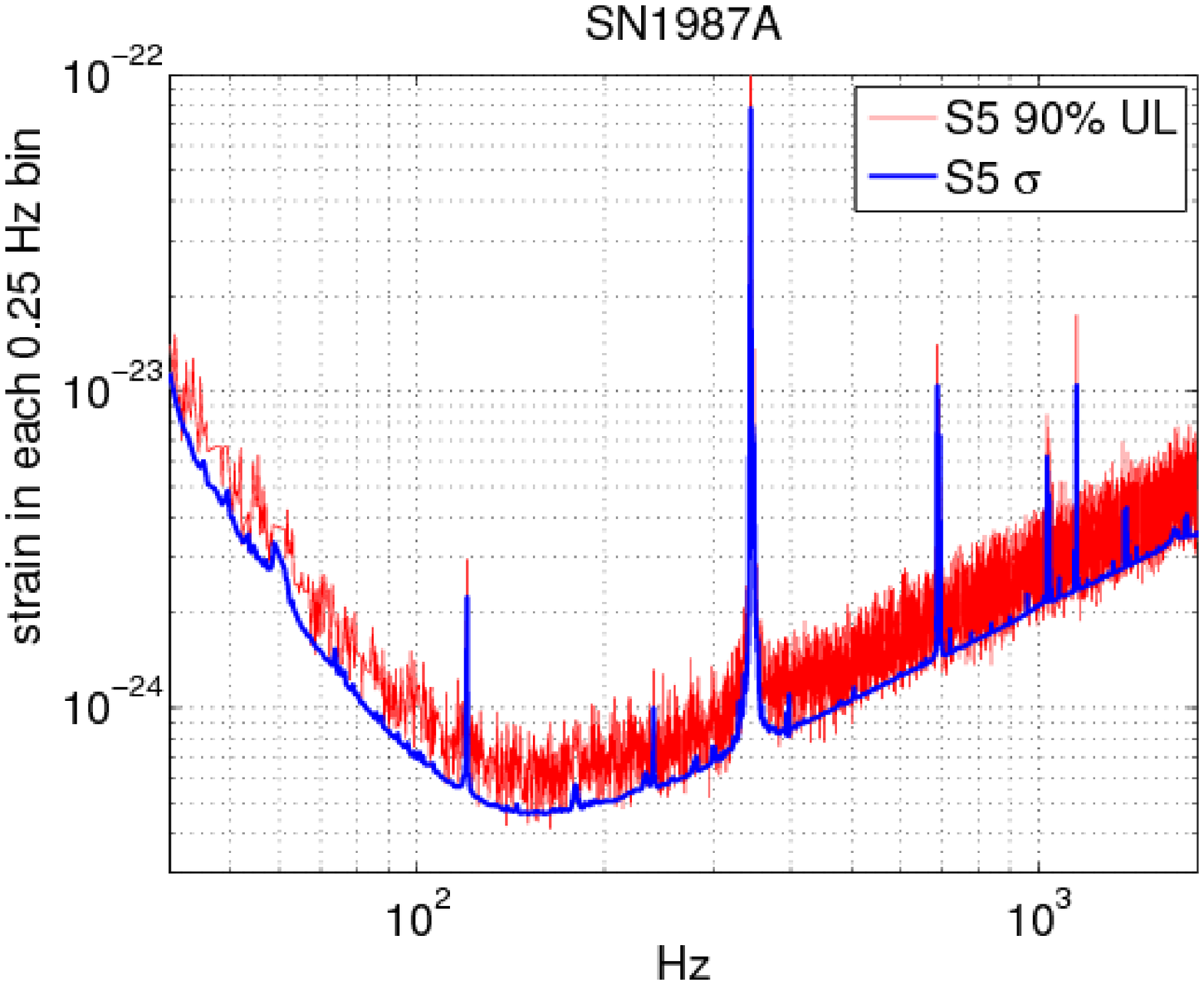,width=2.5in}
  \end{tabular}
  \caption{
    Radiometer 90\% upper limits on RMS strain in each $\unit[0.25]{Hz}$ wide bin as a function of frequency for Sco X-1 (top-left), the Galactic Center (top-right) and SN1987A (bottom).  
    The large spikes correspond to harmonics of the 60~Hz power mains, calibration
    lines and suspension-wire resonances.
    The previous S4 upper limits for Sco X-1 \cite{radiometer} are also plotted in the 
    top-left panel, illustrating the improvement in the upper limits obtained by using 
    the LIGO S5 data.
    This is the first radiometer measurement of the Galactic Center and SN1987A.
    \label{fig:ScoX1}
    }
\end{figure*}

When comparing the SHD analysis with $\beta=0$ and the radiometer upper limits 
obtained using the same spectrum,
it is important to note that these maps have different units.
The radiometer map has units of $\unit[]{\strain^2 Hz^{-1}}$ 
because the radiometer analysis effectively integrates the power 
from a GW point source over solid angle.
The SHD maps, on the other hand, have units of $\unit[]{\strain^2 Hz^{-1} sr^{-1}}$.
If we scale the SHD limit maps by the typical diffraction limited resolution  ($\bar{A}\approx\unit[0.1]{sr}$),
then the limits are more comparable.
The radiometer algorithm limits are lower 
(by a factor of $\lesssim2$) because it requires a stronger 
assumption about the signal model (a single point source),
whereas the SHD algorithm is model-independent.

Figure~\ref{fig:Cls} show 90\% CL upper limits on the $C_l$.
Since the $\hat{\cal P}_{lm}$ have units of strain power
($\unit[]{\strain^2 Hz^{-1}sr^{-1}}$),
the $C_l$ have the somewhat unusual units of
$\unit[]{\strain^4 Hz^{-2}sr^{-2}}$.

{\em Targeted searches:}
Sco X-1 is a nearby ($\unit[2.8]{kpc}$) low-mass X-ray binary likely to include a neutron star spun up through accretion.
Its spin frequency is unknown.
It has been suggested that this accretion torque is balanced by GW emission~\cite{chakrabarty}.
The Doppler line broadening due to the orbital motion is smaller than the chosen $\delta f = \unit[0.25]{Hz}$ bin width for frequencies below $\approx\unit[930]{Hz}$~\cite{Steeghs}.
At higher frequencies, the signal is certain to span two bins.
We determine the maximum value of $\text{SNR}$ 
in the direction of Sco X-1 to be $3.6$ at 
$f=\unit[1770.50]{Hz}$, which has a significance of $p=73\%$ 
given the ${\cal O}(7000)$ independent frequency bins.
Thus in Fig.~\ref{fig:ScoX1} (first panel) we 
present limits on root-mean-square (RMS) 
strain, $h_\text{RMS}(f,\hat\Omega)$, as a 
function of frequency in the direction 
of Sco X-1 $(\text{RA},\text{dec})=(\unit[16.3]{hr},15.6^\circ)$.
These limits improve on the previous best limits by a factor of $\sim5$~\cite{radiometer}.
RMS strain is related to narrow-band GW power via
\begin{equation}
  h_\text{RMS}(f,\hat\Omega) = \left[ \eta(f,\hat\Omega) \delta f \right]^{1/2},
\end{equation}
and is better suited for comparison with searches for periodic GWs, 
which typically constrain the peak strain amplitude, 
$h_0$, marginalized over neutron star parameters $\iota$ and $\psi$ (see, e.g.,~\cite{scox1_s2}).
Our limits on $h_\text{RMS}$ are for a circularly polarized signal from a pulsar whose spin axis is aligned with the line of sight.
Marginalizing over $\iota$ and $\psi$ and converting from RMS to peak amplitude causes the limits to change by a sky-dependent factor of $\approx2.3$~\cite{chris}.
We note that these limits are on the RMS strain in each bin as opposed to the total RMS strain from Sco X-1, which might span as many as two bins.
The frequency axis refers to the observed GW frequency as opposed to the intrinsic GW frequency.

We also look for statistically significant outliers 
associated with the Galactic Center 
$(\text{RA},\text{dec})=(\unit[17.8]{hr},-29^\circ)$ and SN1987A
$(\text{RA},\text{dec})=(\unit[5.6]{hr},-69^\circ)$.
The maximum SNR values are $3.5$ at $f=\unit[203.25]{Hz}$ with $p=85\%$ and $4.3$ at $\unit[1367.25]{Hz}$ with $p=7\%$, respectively.
Limits on RMS strain  are given in the right panel of Fig.~\ref{fig:ScoX1}.

\section{Conclusions}
We performed two directional analyses for persistent GWs:
the radiometer analysis, which is optimized for point sources, 
and the complementary spherical-harmonic decomposition (SHD) algorithm, 
which allows for arbitrary extended sources.
Neither analysis finds evidence of GWs. Thus we 
present upper-limit maps of GW power and also limits on the 
RMS strain from Sco X-1, the Galactic Center and SN1987A.
The radiometer map limits improve on the previous 
best limits~\cite{radiometer} by a factor of 30 in strain power, 
and limits on RMS strain from Sco X-1 constitute 
a factor of 5 improvement in strain over the previous best limits~\cite{radiometer}.
The SHD clean maps represent the first effort to 
look for anisotropic extended sources of GWs.

With the ongoing construction of second-generation 
GW interferometers, we are poised to enter a new era in GW astronomy.
Advanced detectors~\cite{ALIGO,ALIGO2,Virgo,LCGT} are expected to achieve strain sensitivities 
approximately $10$ times lower than initial LIGO, and advances 
in seismic isolation are expected to extend the frequency band 
down from $\unit[40]{Hz}$ to $\unit[10]{Hz}$~\cite{ALIGO}.
By adding additional detectors to our network, 
we expect to reduce degeneracies in the Fisher matrix and improve angular resolution.
These improvements will allow advanced detector 
networks to probe plausible models of astrophysical 
stochastic foregrounds and some cosmological models such as cosmic strings.

\begin{acknowledgments}
The authors gratefully acknowledge the support of the United States
National Science Foundation for the construction and operation of the
LIGO Laboratory, the Science and Technology Facilities Council of the
United Kingdom, the Max-Planck-Society, and the State of
Niedersachsen/Germany for support of the construction and operation of
the GEO600 detector, and the Italian Istituto Nazionale di Fisica
Nucleare and the French Centre National de la Recherche Scientifique
for the construction and operation of the Virgo detector. The authors
also gratefully acknowledge the support of the research by these
agencies and by the Australian Research Council, 
the International Science Linkages program of the Commonwealth of Australia,
the Council of Scientific and Industrial Research of India, 
the Istituto Nazionale di Fisica Nucleare of Italy, 
the Spanish Ministerio de Educaci\'on y Ciencia, 
the Conselleria d'Economia Hisenda i Innovaci\'o of the
Govern de les Illes Balears, the Foundation for Fundamental Research
on Matter supported by the Netherlands Organisation for Scientific Research, 
the Polish Ministry of Science and Higher Education, the FOCUS
Programme of Foundation for Polish Science,
the Royal Society, the Scottish Funding Council, the
Scottish Universities Physics Alliance, The National Aeronautics and
Space Administration, the Carnegie Trust, the Leverhulme Trust, the
David and Lucile Packard Foundation, the Research Corporation, and
the Alfred P. Sloan Foundation.
This is LIGO document \#P1000031.
\end{acknowledgments}

\bibliography{sph_results}

\end{document}